# A LightGBM based Forecasting of Dominant Wave Periods in Oceanic Waters

Pujan Pokhrel
Canizaro Livingston Gulf States
Center for Environmental Informatics
University of New Orleans
New Orleans, Louisiana
ppokhrel@uno.edu

## ABSTRACT

In this paper, we propose to forecast dominant wave periods in oceanic waters using the Light Gradient Boosting (LightGBM) machine learning algorithm. First, we apply various data filtering methods to the dataset collected from CDIP buoys, which allow us to obtain a high-quality dataset for training and validating the proposed models. We extract various wave-based features such as wave heights, periods, skewness, kurtosis, etc., and atmospheric features such as humidity, pressure, and air temperature for the buoys. Subsequently, we train Extra Trees as the base model to compare with the proposed LightGBM through a hv-block cross-validation scheme for up to 30 days ahead. The test or future data was carefully separated from the training dataset to ensure accurate performance evaluations of the models. LightGBM has the $R^2$ score of 0.94, 0.94, and 0.94 for 1-day ahead, 15-day ahead, and 30-day ahead prediction. Similarly, Extra Trees (ET) has an $R^2$ score of 0.88, 0.86, and 0.85 for 1-day ahead, 15-day ahead, and 30 days ahead prediction. Since the LightGBM algorithm outperforms ET for all the windows tested, it is taken as the final algorithm. Note that the performances of both methods do not decrease significantly as the forecast horizon increases.

For short-term prediction, we perform forecasts up to 24 hours ahead in 3-hour intervals, with each window of 30 minutes. None of the existing methods forecast the dominant wave period for 30 minutes window. Similarly, for long-term prediction, we perform forecasts for each day to measure the performance of both machine learning algorithms for up to 30 days ahead, which has also not been explored before. Similarly, the proposed method outperforms the state-of-the-art numerical approaches based on the test dataset. For 1 day ahead prediction, the proposed algorithm has SI, Bias, CC, and RMSE of 0.09, 0.00, 0.97, and 1.78 compared to 0.268, 0.40, 0.63, and 2.18 for a leading numerical forecasting model, namely the European Centre for Medium-range Weather Forecasts (ECMWF).

## 1 Introduction

Prediction of ocean wave parameters remains one of the most important problems of classical physics [1]. The prediction of ocean waves and their properties is important for offshore engineering projects like the exploitation of marine renewable energy, harbor construction, marine operations, marine, and navy marine forces operations, etc. [2-7]. Since the physical process of wave generation is uncertain, nonlinear, and non-stationary, dominant wave period prediction remains difficult [8]. The dominant wave period is defined as the wave period associated with the highest energetic waves at a specific point or area in the total wave spectrum and is always the swell period or wind-wave period. The main approaches for predicting dominant wave periods fall into two categories: energy balance equation-based numerical approaches and statistical approaches.

The most widely used numerical approaches are Wave Analysis Model (WAM) [9], Simulating Waves Nearshore (SWAN) [10], and WAVEWATCH [11-13]. Numerical models predict waves over the large spatial and temporal domain and are beneficial for natural disaster planning, marine activity planning, etc. Although numerical models are considered accurate for the small-time domain, they are computationally costly and time-consuming, which limits their use for feedback control operations [14]. Furthermore, they have low generalization ability for different domains, and the models must be run periodically, resetting the boundary conditions for different regions [8].

To deal with these problems, various data-driven and statistical methods have been proposed. These statistical models can generalize to a wide range of conditions once the statistics are figured out, and they run with less computational and time complexity than their numerical counterparts [8, 14].

The most widely used statistical technique for wave height and period prediction is linear regression [15], ARIMA [16], tree-based methods (bagging and boosting) [17, 18], kernel-based approaches [19, 20], and neural networks [21-24]. The classical empirical-based models, such as Auto Regressive Moving Average (ARMA), are well established, but they cannot easily capture non-stationarity and nonlinearity in the data series [14, 25]. Although numerical models have been proven to be effective in wave height prediction over a large spatial and temporal change, they have higher computational and time costs, especially for the calculations of a higher resolution grid in the nearshore zones where the seabed topography is intricate [14, 25].

Rao and Mandal [21] hindcasted wave heights and periods from the cyclone-generated wind fields using a two-input configuration of Neural Networks. Mahjoobi and Shahidi [2, 19, 26, 27] used feedforward networks with three layers to forecast wave direction, significant wave height, and dominant wave periods. Sensitivity analysis showed that wind speed and direction are the most important parameters for wave hindcasting. Wu *et al.* also forecasted ocean wave parameters like significant wave heights, dominant wave periods, and mean wave periods in the Northern Sea. The datasets used in the paper were derived from ECMWF CERA-20C datasets which are derived from the numerical equations. The temporal resolution of the dataset is 3 hours, and the spatial resolution is $0.125°$ (about 13.7 km). While the $R^2$ score is 0.9196 for a 3-hour ahead prediction, it decreases significantly to 0.7125 for a 12-hours ahead prediction. From Section 4, we can see that the $R^2$ score for the ECMWF model is 0.63 for 24 hours ahead prediction, and thus, training a machine learning model on the dataset generated by the ECMWF model means that the model is learning the numerical equations and not the actual dominant wave periods that is observed in nature. Moreover, there have been many attempts to forecast wave parameters, tide, and short-term operational water levels [2-5, 7, 21, 23, 25, 28-32].

The earlier machine learning models do not employ all the relevant features, including third and fourth-order nonlinearities (skewness and kurtosis), humidity, air temperatures, to model peak wave period in oceans. They also do not forecast till 30 days ahead which is the setup of our experiment. Moreover, since they do not employ various methods for data filtering, the training data has many outliers, which mainly occur due to factors like buoys getting carried with the waves, biofouling, etc. They also do not perform time-based hv block cross-validation and do not take seasonality and time variables (hour, day, month, and year) in the dataset. These factors reduce the predictability of wave parameters and reduce the accuracy of data-driven approaches. Moreover, none of the earlier studies have focused on predicting dominant wave periods for both short- and long-term predictions. While the short-term prediction is important for avoiding instantaneous large waves in the oceans, long-term prediction is important for ship routing and other environmental planning operations.

This study proposes a machine-learning-based approach to forecast dominant wave periods in oceans using features derived from the buoys like previous significant wave heights, periods, wave direction, directional spread, wave periods (mean/zero-crossing/dominant), and the features derived from NOAA like wind properties, humidity, temperature, and pressure. We perform both short-term (0-24 hrs prediction) and long-term predictions (1-30 days) to forecast dominant wave periods in oceanic waters. Note that none of the existing methods can forecast for 30-minute windows, and also, none of them can forecast for up to 30 days ahead. We use the CDIP buoy dataset because it provides an accurate point-based measurement that provides high-resolution data for machine learning methods. Likewise, NOAA gives real-time wave nowcast and forecasts coupled with buoy data for accurate operational prediction. The data filtering procedures remove erroneous buoy data and help create an accurate and robust model for prediction. Moreover, tree-based algorithms like LightGBM and Extra Trees, along with time-based hv block cross-validation, allow the machine learning algorithms to be explainable while capturing various time dependencies on the data and giving superior performance compared to numerical approaches.

The main contributions of this paper are as follows.

1) The proposed method outperforms the state-of-the-art numerical approaches for dominant wave period prediction.

2) We use the dataset derived from actual CDIP buoys that have a temporal resolution of 30 minutes and give point measurements, which provide accurate information about the ocean properties.

3) The data filtering techniques used in this paper remove the erroneous data points from the training dataset so that the model only learns relevant relationships to make predictions.

4) The use of tree-based machine learning algorithms like LightGBM and Extra Trees allows us to get the feature importance and learn the decision process behind individual predictions.

5) The use of features like skewness, directional spread, power spectral density, and kurtosis helps capture various nonlinearities and help the models make accurate predictions.

## 2. Methodology

### 2.1 Dataset

The buoy dataset used in this paper is derived from CDIP [33]. Likewise, the dataset for pressure, wind components, specific and relative humidity is derived from NOAA Reanalysis II [34]. The dataset covers the period 1948 to the present. Note that the CDIP buoy dataset has a temporal resolution of 30 minutes, whereas all the other datasets have a temporal resolution of 6 hours. Likewise, NOAA data covers the whole planet, though in coarse resolution. CDIP data is fine-grained and more accurate, but buoys are not available everywhere around the Earth.

## 2.2 Features

The wave parameters are derived from buoys and NOAA datasets. The procedure for derivation of wave features has been discussed in [35], and the procedure for the derivation/access of wind variables, pressure, humidity, and air temperature is discussed at [34]. The variables used in this paper include:

**Table 1: Properties of various features used in this dataset.**

| Variable | Min | Max | Average | Std |
|---|---|---|---|---|
| Depth | 7.5 | 1910 | 270.26 | 492.45 |
| Directional Spread | 23.66 | 71.12 | 44.80 | 6.17 |
| Significant wave height | 0.02 | 15.42 | 0.99 | 0.50 |
| Zero crossing period | 1.84 | 15.38 | 5.24 | 1.55 |
| Dominant wave period | 1.59 | 25 | 10.14 | 4.21 |
| Power Spectral Density | 0.003 | 107.47 | 1.5 | 0.49 |
| Mean wave period | 1.85 | 16.92 | 6.0 | 1.92 |
| Wave direction | 0.25 | 359 | 193.42 | 83.16 |
| Skewness | -15.35 | 11.37 | -0.35 | 1.18 |
| Kurtosis | 1.08 | 5.99 | 3.28 | 1.19 |
| uwind | -18.30 | 24.70 | 0.922 | 5.899 |
| vwind | 20.70 | 17.80 | 1.69 | 4.55 |
| Pressure (omega) | -0.286 | 0.260 | -0.001 | .051 |
| Relative humidity | 4.0 | 100.0 | 80.21 | 11.02 |
| Specific Humidity | 0.0003 | 0.021 | 0.010 | 0.003 |
| Air temp | 269.0 | 319.7 | 294.31 | 6.34 |

The other variables included are latitude, longitude, depth, time of the day, day of the month, and month of the year.

## 2.3 Machine Learning Methods

We use two machine learning methods in this study: Extra Trees (ET) [36] and Light Gradient Boosting Machine (LightGBM) [37]. Both are tree-based approaches, where the former performs bagging, whereas the latter performs boosting. Bagging and boosting are commonly used ensemble methods for prediction purposes. The ET algorithm builds an ensemble of unpruned regression trees according to the classical top-down procedure. In ET, the nodes are split by choosing the cut-points at random and uses the whole learning sample to grow trees. While the splits are done at random in the case of ET, LightGBM utilizes Gradient-based One Side Sampling and Exclusive Feature Bundling to down-sample the training data and features, thus decreasing training time. Both methods give interpretability in the sense that the feature importance is given by default, and the decision process for individual decisions is also easy to figure out. In summary, the use of these machine learning algorithms is motivated by:

1) Both algorithms provide interpretability in the sense that feature importance is provided by default, and the decision process is easy to follow.

2) Both are ensemble methods, which perform bagging and boosting on the data. Bagging and boosting are common statistical methods to improve accuracy.

3) The training time for both algorithms is low compared to other methods.

## 2.4 Data Filtering Procedures

Field measurement of waves is subject to various errors that must be removed to obtain a high-quality and reliable dataset. Some examples of the errors include experimental error, buoys getting carried away with the waves, underestimating wave heights, biofouling, etc. Thus, a stringent Quality Control (QC) procedure is required to obtain a good dataset in which various machine learning methods can be trained.

Therefore, while most of the earlier works on waves prediction train their machine learning models on the NOAA buoy dataset [2, 21, 26, 27, 38], we use the data obtained from CDIP buoys, which uses a rigorous QC Procedure [39]. NOAA performs basic checks on the range of various parameters [40, 41] whereas CDIP performs robust tests that provide a high-quality dataset for training and validation [42]. CDIP buoys perform various tests on the time series, spectral, and parameter values to obtain a high-quality dataset [42]. While the NOAA buoys dataset is provided in hourly intervals, the buoy data from CDIP is provided in 30-minute intervals for time-averaged spectral data. These factors motivate the use of CDIP data for training the machine learning models in our paper.

The CDIP buoys automatically flag all questionable, bad, or missing data points in the same domain as the vertical displacement. CDIP then runs a robust shore-side QC procedure [39] to identify the erroneous data points. For the erroneous data that was not identified by the buoy or CDIP QC procedure, a series of filters were employed to screen and remove the outliers. The filtering procedure is based on work by Christou and Ewans [43] and has been used widely throughout the literature [44-47].

1) Individual waves with zero-crossing wave period > 25s.
2) Rate of change of surface elevation Sy exceeded by a factor of 2.

$$Sy = \frac{2\pi\sigma}{Tz}\sqrt{2lnN_z} \qquad (1)$$

where $\sigma$ is the standard deviation of the surface elevation $\eta$ and $N_z$ is the number of zero up crossing periods (Tz).

3) Absolute crest or trough elevation greater than 5 times the standard deviation of the 30-min water surface elevation.
4) A single zero-crossing containing > 2304 points.
5) Wave crest elevation $\eta_c > 1.5 H_s$.
6) Horizontal buoy excursion $\triangle x, \triangle y$ where $\triangle x > 1.8 H_s$ or $\triangle y > 1.8 H_s$.
7) Kurtosis values less than 2 or greater than 8.
8) Wave steepness greater than 0.142.

## 2.5 Cross-validation

We use the hv-block cross-validation procedure [48], first proposed by Racine and used for stationary time series data. The procedure is like the KFold validation, except that there is no random shuffling of the observations. Thus, it renders K blocks of contiguous observations in their natural order. Afterward, while testing the models, adjacent observations between the training and test sets are removed to create a gap between the two sets and increase independence among the observations. The number of folds used in this paper is 10. The general procedure for hv-block cross-validation is illustrated in Figure 1.

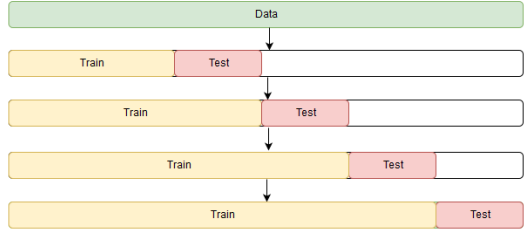

**Figure 1: Setup for hv-blocked cross-validation**

## 2.6 Evaluation Metrics

To measure the performance of our model and to compare the results with other methods, we employ various metrics like Root Mean Squared Error (RMSE), Mean Absolute Error (MAE), Variance, $R^2$ Score, Scatter Index (SI), Correlation Coefficient (CC), Bias, and Hanna and Heinold (HH) Indicator. Note that RMSE, Bias and MAE are measured in $meters$, Variance in $meters^2$ and SI, $R^2$ Score, CC, and HH are nondimensional.

**Table 2: Evaluation Metrics and their calculations**

| Name | Mathematical Formula |
|------|---------------------|
| RMSE | $RMSE = \sqrt{\dfrac{\sum_{i=1}^{N}(x_i - \hat{x}_i)^2}{N}}$ |
| MAE | $MAE = \dfrac{1}{N}\sum_{i=1}^{N}|x_i - \hat{x}_i|$ |
| Variance | $Variance = \dfrac{\sum_{i=1}^{N}(x_i - x_m)^2}{N}$ |
| $R^2$ score | $R2 = 1 - \dfrac{\sum_{i=1}^{N}(x_i - \hat{x}_i)^2}{\sum_{i=1}^{N}(x_i - x_m)^2}$ |
| SI | $SI = \dfrac{RMSE}{\frac{1}{n}\sum_{i=1}^{N}x_i}$ |
| CC | $CC = \dfrac{\sum_{i=1}^{N}(x_i - x_m)(\hat{x}_i - \hat{x}_m)}{\sqrt{\sum_{i=1}^{N}(x_i - x_m)^2 \sum_{i=1}^{N}(\hat{x}_i - \hat{x}_m)^2}}$ |
| Bias | $Bias = \dfrac{1}{N}\sum_{i=1}^{N}(x_i - \hat{x}_i)$ |
| HH | $HH = \sqrt{\dfrac{\sum_{i=1}^{N}(x_i - \hat{x}_i)^2}{\sum_{i=1}^{N}x_i \hat{x}_i}}$ |

In the preceding table, $i$, $x_i$, $\hat{x}_i$, $x_m$, $\hat{x}_m$, and $N$ refer to the position, measured value, predicted value, mean of actual values, mean of predicted values, and the number of elements, respectively.

Ten-fold hv-blocked cross-validation was used for training, and then the optimum parameters were identified. After calculating the best parameters, we train the data on the training set and predict the data using the test set. Note that each window contains 30-min data, and thus, for day-1 forecasting, the algorithm forecasts 48 steps and so on.

## 3 Results

### 3.1 Short term predictions

For the short-term prediction, we predict the dominant wave periods in 3-hour intervals. Note that while 3 hourly intervals are taken, the window size of both input and forecast period is 30 minutes. Thus, every 3 hourly prediction adds 6 steps to our proposed method. Figures (2-9), Table 3, and Table 4 show the performance of LightGBM and ET for predictions up to 24 hours ahead in 3-hour intervals.

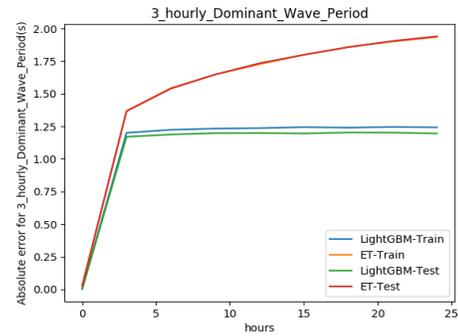

**Figure 2: Absolute Error for dominant wave periods for 3-hour interval forecasts.**

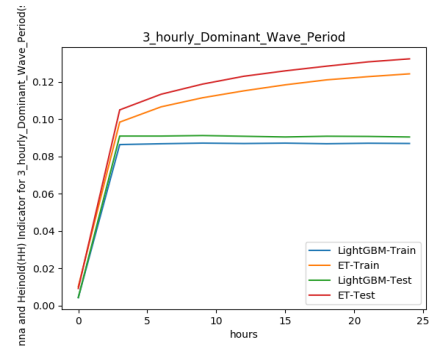

**Figure 3: HH Indicator for dominant wave periods for 3-hour interval forecasts.**

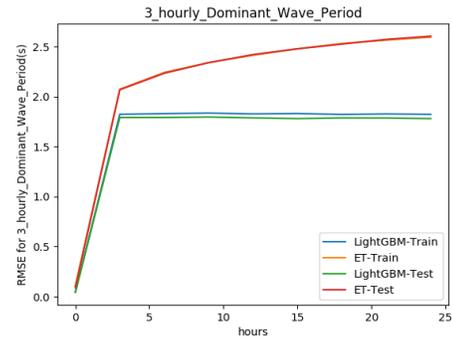

**Figure 4: Root Mean Squared Error for dominant wave periods for 3-hour interval forecasts.**

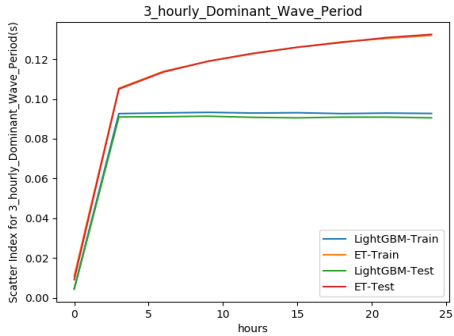

**Figure 5: Scatter Index (SI) for dominant wave periods for 3-hour interval forecasts.**

**Table 3: MAE, RMSE, HH, and SI for machine learning algorithms for short-term predictions.**

| Algorithm | Dataset | Metric | 3 hr | 15 hr | 24 hr |
|---|---|---|---|---|---|
| LightGBM | Train | MAE | 1.20 | 1.24 | 1.25 |
| | | RMSE | 1.83 | 1.83 | 1.83 |
| | | HH | 0.08 | 0.09 | 0.09 |
| | | SI | 0.09 | 0.09 | 0.09 |
| | Test | MAE | 1.16 | 1.20 | 1.20 |
| | | RMSE | 1.79 | 1.78 | 1.78 |
| | | HH | 0.09 | 0.09 | 0.09 |
| | | SI | 0.09 | 0.09 | 0.09 |
| ET | Train | MAE | 1.37 | 1.80 | 1.93 |
| | | RMSE | 2.07 | 2.50 | 2.60 |
| | | HH | 0.09 | 0.11 | 0.13 |
| | | SI | 0.10 | 0.13 | 0.13 |
| | Test | MAE | 1.37 | 1.80 | 1.94 |
| | | RMSE | 2.07 | 2.48 | 2.61 |
| | | HH | 0.10 | 0.13 | 0.13 |
| | | SI | 0.10 | 0.13 | 0.13 |

In Figures (2-5) and Table 2, we can see that the performance of Extra Trees decreases rapidly from 0 to 3 hours but then decreases slowly for MAE, RMSE, HH, and SI. Likewise, for LightGBM, the performance decreases sharply from 0 to 3 hours but then remains fairly consistent afterward with small fluctuations. The interesting thing to note here is that while the performance of ET continues to decrease till 24 hours, the performance of LightGBM does not decrease similarly.

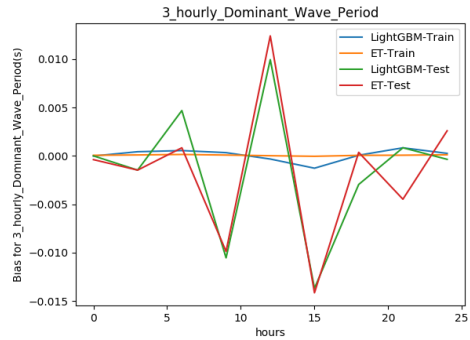

**Figure 6: Bias Error for dominant wave periods for 3-hour interval forecasts.**

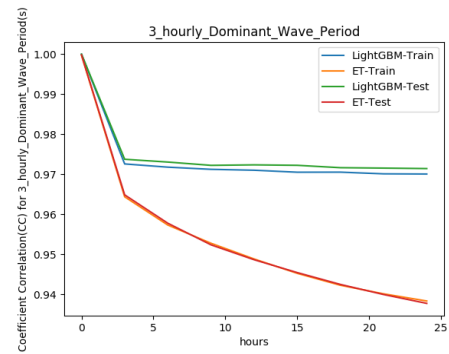

**Figure 7: Coefficient Correlation (CC) for dominant wave periods for 3-hour interval forecasts.**

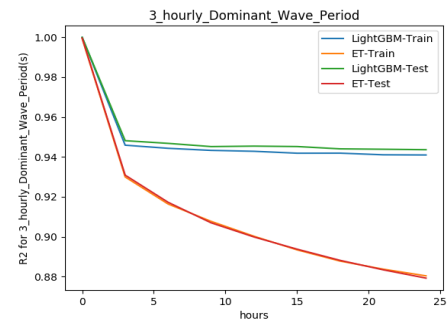

**Figure 8: Adjusted $R^2$ for dominant wave periods for 3-hour interval forecasts**

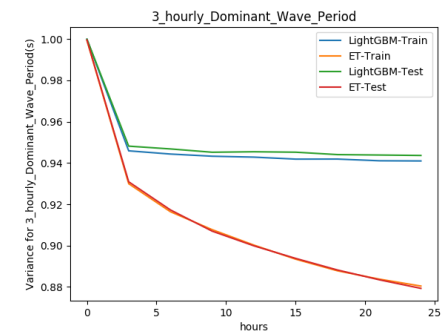

**Figure 9: Variance Error for dominant wave periods for 3-hour interval forecasts.**

**Table 4: Bias, CC, $R^2$, and Variance for machine learning algorithms for short-term predictions.**

| Method | Dataset | Metric | 3 hr | 15 hr | 24 hr |
|--------|---------|--------|------|-------|-------|
| LightGBM | Train | Bias | 0.00 | 0.00 | 0.00 |
| | | CC | 0.98 | 0.97 | 0.97 |
| | | $R^2$ | 0.95 | 0.94 | 0.94 |
| | | Variance | 0.95 | 0.94 | 0.94 |
| | Test | Bias | 0.00 | 0.01 | 0.00 |
| | | CC | 0.97 | 0.97 | 0.97 |
| | | $R^2$ | 0.95 | 0.95 | 0.94 |
| | | Variance | 0.95 | 0.95 | 0.94 |
| ET | Train | Bias | 0.00 | 0.00 | 0.00 |
| | | CC | 0.96 | 0.95 | 0.94 |
| | | $R^2$ | 0.93 | 0.89 | 0.88 |
| | | Variance | 0.93 | 0.89 | 0.88 |
| | Test | Bias | 0.00 | 0.01 | 0.00 |
| | | CC | 0.97 | 0.95 | 0.94 |
| | | $R^2$ | 0.93 | 0.89 | 0.88 |
| | | Variance | 0.93 | 0.89 | 0.88 |

From Figures (6-9) and Table 3, we observe that the performance of the machine learning algorithms decreases as the time steps for forecasting increase. Specifically, for both algorithms, Coefficient Correlation, Variance, Adjusted $R^2$ decrease rapidly for predictions till 15 hours but decrease slowly afterward. While the ET algorithm continues to decrease after 3 hours, the LightGBM algorithm does not decrease significantly after 3 hours. Moreover, for bias, the error increases till 15 hours but then decreases afterward for both algorithms. Since the bias values are within $\pm 0.002$, they are very small and considered insignificant.

## 3.2 Long term predictions

In this section, we present the results of the proposed method on longer time frames. The interval between forecast periods is 24 hrs, i.e., for day-1 ahead prediction, we forecast $2 \times 24 = 48$ steps; for day-2, we forecast $2 \times 2 \times 24 = 96$ steps, and so on. Figures (10-17) and Table 5 and Table 6 show the performance of LightGBM and ET on the dataset for dominant wave period prediction for days ahead prediction.

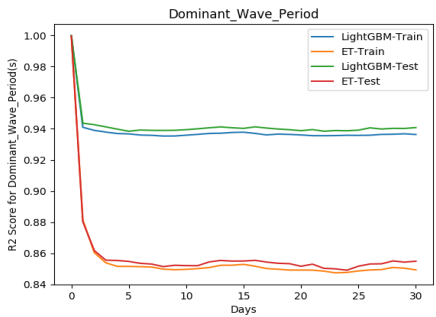

**Figure 10: Adjusted $R^2$ score for the significant wave heights for daily predictions.**

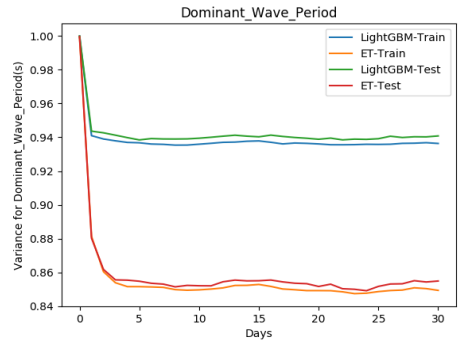

**Figure 11: Variance for dominant wave periods for daily predictions.**

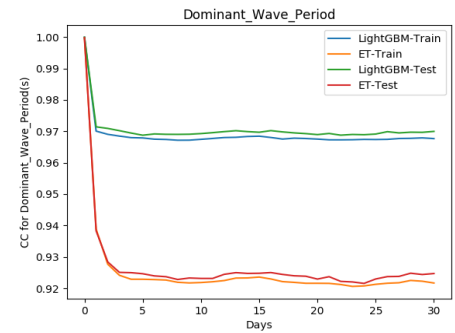

**Figure 12: Correlation Coefficient (CC) for dominant wave periods daily predictions.**

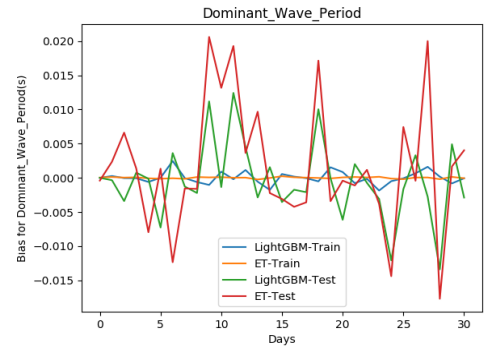

**Figure 13: Bias score for significant wave heights for daily predictions.**

**Table 5: Bias, CC, $R^2$, and Variance for machine learning algorithms for long-term prediction**

| Method | Dataset | Metric | 1 day | 15 days | 30 days |
|--------|---------|--------|-------|---------|---------|
| LightGBM | Train | Bias | 0.00 | 0.00 | 0.00 |
| | | CC | 0.97 | 0.97 | 0.97 |
| | | $R^2$ | 0.94 | 0.94 | 0.94 |
| | | Variance | 0.94 | 0.94 | 0.94 |
| | Test | Bias | 0.00 | 0.00 | 0.00 |
| | | CC | 0.97 | 0.97 | 0.97 |
| | | $R^2$ | 0.94 | 0.94 | 0.94 |
| | | Variance | 0.95 | 0.94 | 0.94 |
| ET | Train | Bias | 0.00 | 0.00 | 0.00 |
| | | CC | 0.94 | 0.93 | 0.92 |
| | | $R^2$ | 0.88 | 0.86 | 0.85 |
| | | Variance | 0.88 | 0.86 | 0.85 |
| | Test | Bias | 0.00 | 0.00 | 0.00 |
| | | CC | 0.94 | 0.94 | 0.92 |
| | | $R^2$ | 0.88 | 0.86 | 0.85 |
| | | Variance | 0.88 | 0.86 | 0.85 |

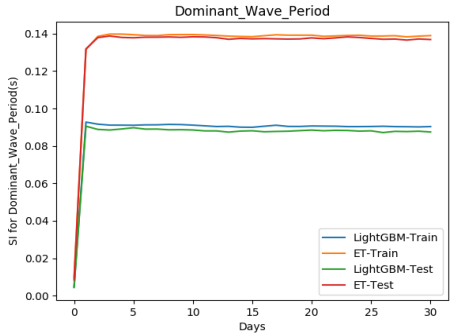

**Figure 14: Scatter Index (SI) for significant wave heights for daily predictions.**

From Figures (10–13) and Table 5, we note that $R^2$, Variance, Coefficient Correlation remain fairly constant with some fluctuations even though the forecast range increases. Specifically, for the LightGBM algorithm, the performance does not decrease significantly after a 3-hour ahead prediction. However, in the case of Extra Trees, it continues to decrease till 3 days ahead prediction, even though the drop in performance is less significant when compared to 1 day ahead prediction, and remains fairly constant afterward. We attribute the small fluctuations in performance afterward to the outliers in the dataset and the data being noisy. Moreover, while the performance of the algorithms decreases rapidly from day-0 to day-1 forecasting, it does not decrease significantly afterward. Likewise, while the value of bias fluctuates, it remains within $\pm 0.020$ throughout the forecasting periods tested.

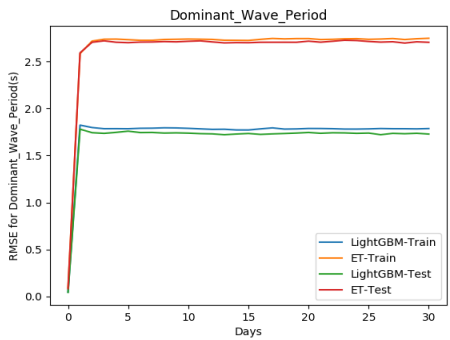

**Figure 15: Root mean squared error (RMSE) score for the dominant wave periods for various forecast ranges.**

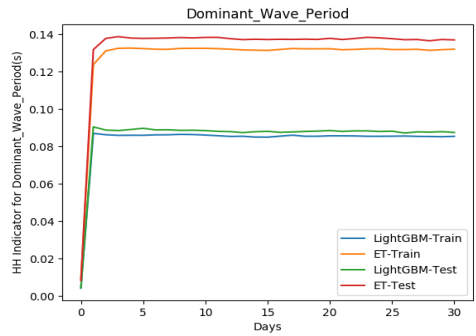

**Figure 16: Hanna and Heinold (HH) Indicator for significant wave heights for daily predictions.**

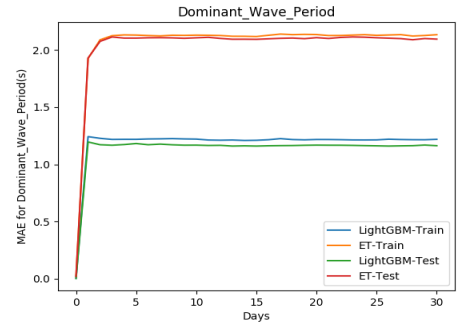

**Figure 17: Mean Absolute Error (MAE) for the significant wave heights for daily predictions.**

**Table 6: MAE, RMSE, HH, and SI for machine learning algorithms for long-term forecasting.**

| Method | Dataset | Metric | 1 day | 15 days | 30 days |
|---|---|---|---|---|---|
| LightGBM | Train | MAE | 1.24 | 1.21 | 1.22 |
| | | RMSE | 1.82 | 1.78 | 1.79 |
| | | HH | 0.09 | 0.08 | 0.09 |
| | | SI | 0.09 | 0.09 | 0.09 |
| | Test | MAE | 1.20 | 1.16 | 1.17 |
| | | RMSE | 1.78 | 1.74 | 1.73 |
| | | HH | 0.09 | 0.09 | 0.09 |
| | | SI | 0.09 | 0.09 | 0.09 |
| ET | Train | MAE | 1.92 | 2.12 | 2.14 |
| | | RMSE | 2.59 | 2.73 | 2.75 |
| | | HH | 0.12 | 0.13 | 0.14 |
| | | SI | 0.13 | 0.14 | 0.14 |
| | Test | MAE | 1.93 | 2.09 | 2.10 |
| | | RMSE | 2.59 | 2.70 | .71 |
| | | HH | 0.13 | 0.14 | 0.14 |
| | | SI | 0.13 | 0.14 | 0.14 |

From Figure 14 to Figure 17 and Table 6, we can see that the performance of both algorithms LightGBM, and Extra Trees remains constant with some fluctuations while the forecast window increases from 1 day to 30 days. The values of HH indicator, MAE, RMSE, and SI do not increase significantly after 1 day ahead prediction. Moreover, bias shows some fluctuations from day-7 to day-8, day-16 to day-17, and so on. But, since the values for bias errors are very small (less than 0.015), the fluctuations are not significant.

## 4 COMPARISON OF DOMINANT WAVE PERIOD PREDICTION WITH METHODS OF OTHER WEATHER AGENCIES

In this section, we compare our method with other commonly used methods for 1 day ahead forecasting. Note that since Bidlot et al. [49] tested the other methods using buoy data from June to August 2007, we utilize data from January 2001 to June 2007 as a training dataset and June to August 2007 as the test dataset. The models used in the comparison are European Centre for Medium-range Weather Forecasts (ECMWF) [49-51], Met Office (MetO) [31], Fleet Numerical Meteorology and Oceanography Center (FNMOC), National Centers for Environmental Prediction (NCEP) [52-55], Bureau of Meteorology (BoM) [49, 56-58], and Service Hydrographique et Oceanographique de la Marine (SHOM) [59]. While ECMWF outperforms all the other numerical approaches for the test dataset in Table 7, the methods compared are from the national weather centers of various countries, which are optimized for a specific country/region. All global methods we compare our approach with are taking the data from (0-12 hours) to predict up to 24 hours ahead, which is only one step forward prediction.

In contrast, we perform 48 steps ahead prediction while only taking the wave information from (0-0.5 hours). Even though we take the individual prediction window of 30 minutes, which is less stationary than the 12-hour window, our method performs significantly better. It is important to point out that we only compare our approach to numerical methods since earlier machine learning methods were shown to underperform numerical approaches. The results obtained can be summarized in Table 3.

**Table 7: Comparison of the proposed method with other state-of-the-art methods.**

| Method | SI | Bias(s) | CC | RMSE(s) |
|---|---|---|---|---|
| ECMWF | 0.268 | 0.40 | 0.63 | 2.18 |
| MetO | 0.448 | 1.65 | 0.40 | 3.94 |
| FNMOC | 0.315 | -0.21 | 0.54 | 2.53 |
| NCEP | 0.245 | -0.66 | 0.65 | 2.06 |
| BoM | 0.401 | 0.680 | 0.34 | 3.27 |
| SHOM | 0.369 | 1.14 | 0.51 | 3.16 |
| Proposed Method | **0.09** | **0.00** | **0.97** | **1.78** |

In the preceding table, the best values are highlighted in bold. Note that our proposed method has a very low Scatter Index (SI), bias, and RMSE value. Likewise, the Correlation Coefficient (CC) of the proposed method is very high compared to the other methods, suggesting that the proposed method outperforms existing state-of-the-art methods.

## 5 DISCUSSION

The results section shows that, in general, the performance of both algorithms, LightGBM and ET, decreases at first, from day 0 to day 1 but then remains fairly consistent with small fluctuations. Moreover, one interesting outcome of this experiment is that the predictability of the waves does not decline significantly even when the forecast horizon increases (up to 30 days). The values of SI remain around 0.17, CC around 0.97, Variance around 0.88, $R^2$ around 0.94, RMSE around 1.82, and MAE around 1.25. These results show that for longer forecasts periods, the performance is significant and can be used for forecasting dominant wave periods in oceanic waters.

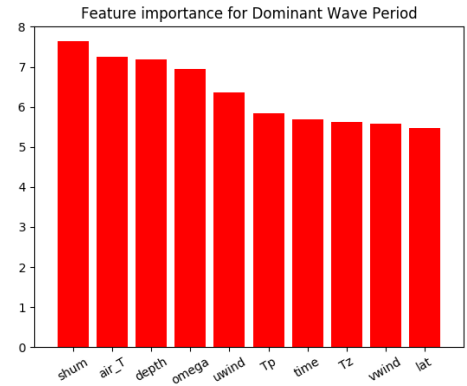

**Figure 18: Feature importance of various features for 6-hour ahead forecasting of dominant wave period.**

In this experiment, LightGBM outperforms Extra Trees (ET) for all the windows tested. The better performance of LightGBM is due to the data containing more bias error than variance error. The bias error arises because the dominant wave period can lie anywhere in the frequency spectrum and often change considerably, without any pattern from one period to the next. LightGBM being built on boosting principle takes advantage of reducing the bias errors.

To characterize how the features are affecting the machine learning algorithms for short-term and long-term predictions, we take the LightGBM algorithm used for 3 hours ahead prediction to plot feature importance for short-term prediction and 1 day ahead prediction for the information about long-term prediction.

Figure 18 shows the importance of the top 10 features in percentage for 3-hours ahead prediction in descending order. Note that since the resolution of the NOAA data is 6 hours, we take 6 hours ahead forecast to measure the feature importance for short-term prediction so that data leakage can be avoided from the features derived from NOAA. In Figure 10, Specific Humidity, Air Temperature, Depth, Pressure, Wind (u-component), Dominant Wave Period, Time of day, Zero crossing period, wind (v-component), and latitude have feature importance of 7.75, 7.43, 7.42, 7.37, 6.91, 5.98, 5.86, 5.72, 5.45, and 5.34, respectively.

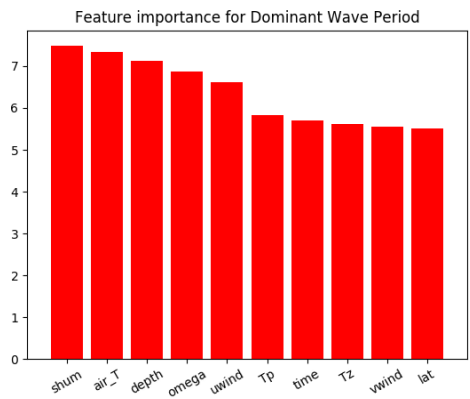

**Figure 19: Feature importance of various features for 24-hour ahead prediction for forecasting of dominant wave period.**

Figure 19 shows the importance of the top 10 features in percentage for 3-hours ahead prediction in descending order. In Figure 10, Specific Humidity, Air Temperature, Depth, Pressure, Wind (u-component), Dominant Wave Period, Time of day, Zero crossing period, wind (v-component), and latitude have feature importance of 7.48, 7.34, 7.13, 6.86, 6.62, 5.82, 5.71, 5.62, 5.55, and 5.50, respectively.

The same features remain in the top 10 for predicting dominant wave periods over 6 hours and 24 hours—however, the contribution of the variables changes as the forecast window increases. For example, for 24 hours ahead prediction, the contribution of wind (v-component) and latitude increased by a small amount compared to 6 hours ahead prediction; however, the contribution of other variables decreases.

We use various data filtering methods to get the most accurate data for training and test purposes. We then use hv-blocked cross-validation to ensure that the data's temporal order is not disturbed while training the model. After that, we compare various machine learning algorithms on the features generated and found that ET, a tree-based machine learning algorithm, works better for forecasting using various wave features derived from the buoys.

From the above results, we observe that, while only taking the data from the 30-min interval, the machine learning methods tested (ET and LightGBM) can forecast up to 30 days (which is the setup of our experiment) without an abrupt decrease in prediction performance. Unlike the numerical methods compared in this paper, which use 12-hour intervals, our proposed algorithm can forecast the significant wave heights in 30-min intervals and still has better performance than the other state-of-the-art methods. The code can be downloaded from:

https://github.com/ppokhrel1/waves.

## 6 Conclusion

This paper explores machine learning approaches for predicting various properties of ocean waves. We use the data from CDIP buoys and NOAA Reanalysis to predict dominant wave periods of ocean waves. The prediction capability has been improved drastically, both in terms of accuracy and forecast horizon, compared to other state-of-the-art approaches due to the high-quality dataset obtained from CDIP buoys, data filtering procedures, and LightGBM machine learning algorithm trained with hv block cross-validation.

We have not only used CDIP buoy data for forecasting dominant wave periods, but we also have used the buoy-based spectral features, temporal information, combined with wind, pressure, humidity, and air temperature to forecast dominant wave periods in oceans. The use of these environmental variables, combined with the machine learning methods trained using hv-block cross-validation, gives improved performance over the other state-of-the-art approaches and allows us to forecast up to 30 days ahead, which is a drastically longer horizon than any other approaches. Similarly, to the best of our knowledge, we are first to try a longer duration, such as forecasting for 30 days. The environmental variables are humidity, air temperature, depth, pressure, wind, previous dominant wave period, time, zero-crossing period, and latitude have the top feature importance. It suggests that information about the environment is crucial for forecasting dominant wave periods. Finally, we believe that our result would serve as a benchmark for future studies on dominant wave prediction involving other environment variables and ranging from small to large scale events for accurate forecasting.